\documentclass[10pt]{article}

\newcommand{\equn}[1]{\begin{equation}\label{#1}}
\newcommand{\eqan}[1]{\begin{eqnarray}\label{#1}}
\newcommand{\eqa}{\begin{eqnarray}}
\newcommand{\equ}{\begin{equation}}
\newcommand{\nuqe}{\end{equation}}
\newcommand{\uqe}{\end{equation}}
\newcommand{\naqe}{\end{eqnarray}}
\newcommand{\aqe}{\end{eqnarray}}

\newcommand{\bk}{\color{black}}

\usepackage{graphicx}
\usepackage{amssymb,amsfonts,amsmath}
\usepackage{color}
\usepackage{enumerate}
\usepackage[utf8]{inputenc}
\usepackage{fancybox}
\usepackage{calc}
\usepackage{esint}
\usepackage{cite}
\usepackage{float}
\usepackage{enumitem}
\usepackage{chemarr}
\usepackage{lineno}
\usepackage{algpseudocode}

\usepackage{amsmath}
\usepackage{amssymb}

\usepackage{cite}

\usepackage{hyperref}

\usepackage{lineno}

\usepackage{microtype}
\DisableLigatures[f]{encoding = *, family = * }

\usepackage[labelfont=bf,labelsep=period,justification=raggedright]{caption}

\makeatletter
\renewcommand{\@biblabel}[1]{\quad#1.}
\makeatother

\date{}

\begin{document}

\begin{center}
{\Large
\textbf{Stochastic Simulation of Biomolecular Networks in Dynamic Environments}
}
\\
Margaritis Voliotis$^{1}$, 
Philipp Thomas$^{2,3}$, 
Ramon Grima$^{3,\ast}$
Clive G. Bowsher$^{1,\dagger}$
\\
{\small \bf{1}} School of Mathematics, University of Bristol, U.K.
\\
{\small \bf{2}} School of Mathematics, University of Edinburgh, U.K.
\\
{\small \bf{3}} School of Biological Sciences, University of Edinburgh, U.K.
\\
{\small $\ast$} E-mail: Ramon.Grima@ed.ac.uk
$\dagger$ E-mail: C.Bowsher@bristol.ac.uk, 
\end{center}

\section*{Abstract}

Simulation of biomolecular networks is now indispensable for studying
biological systems, from small reaction networks to large ensembles of cells.  Here we present a novel approach
for stochastic simulation of networks embedded in
the dynamic environment of the cell and its surroundings. 
{\bk We thus sample trajectories of the stochastic process described by the chemical master equation with time-varying propensities}.
A comparative analysis shows that existing approaches can either fail dramatically, or else can impose impractical computational burdens due
to numerical integration of reaction propensities, especially when cell ensembles are studied. Here we introduce the Extrande method
which, given a simulated time course of dynamic network inputs, provides a conditionally exact
and several orders-of-magnitude faster simulation solution. The new approach makes it feasible to demonstrate---using decision-making by a large population of quorum sensing bacteria---that robustness to 
fluctuations from upstream signaling places strong constraints on the design of networks determining cell fate.
Our approach has the potential to significantly advance both understanding of molecular systems biology and 
design of synthetic circuits.
\vspace{-0.1in}

\section*{Introduction}
Dynamic simulation is an essential and widespread approach for studying biomolecular networks in cell biology \cite{Covert:2012}. However, the computational resources required can quickly become limiting for several reasons. Cellular networks are complex, containing many biomolecular species and reactions. The effects of biochemical stochasticity can be pervasive at the single-cell level \cite{Eldar:2010kk, Thomas2014}, implying that stochastic simulation approaches are often needed. And cells do not live in isolation, which requires simulation on multiple scales, ranging from the single cell to large ensembles of communicating cells \cite{Crampin:2004,Rand:2012}. In these circumstances, parsimonious models of intracellular networks offer dimension reduction \cite{Hartwell:1999ef, Sontag:2007, Thomas2012} and significant advantages \cite{Bialek2013}. However, such models often only provide accurate descriptions when they include the effects of interactions with other fluctuating processes in the cell and of signals arising extracellularly \cite{Bowsher2013,Bowsher2014b,Zechner2014}. 
{\bk While it is straightforward to write a Chemical Master Equation describing the stochastic dynamics of these models, it is usually impenetrable to analysis and one needs to make use of simulation methods.}
The stochastic simulation algorithm (SSA) \cite{Gillespie:1976,Gillespie:1977ww} allows {\bk only} the random timing of reactions in the network model to be taken into account (often known as intrinsic noise), but cannot be used when other processes interacting with the network cause its propensities to fluctuate between reaction occurrences. The SSA assumes constant propensities between reactions (and hence exponentially distributed waiting times). Here we present a new approach relaxing this assumption, called Extrande, for stochastic simulation of a biomolecular network of interest embedded in the dynamic, fluctuating environment of the cell and its surroundings. 

Biological processes that interact with the network or model of interest are sometimes called extrinsic processes \cite{Swain:2002kn}. They often significantly change the stochastic behaviour and dynamics of the network \cite{Hilfinger:2011ev,Bowsher2012}. We briefly give two illustrations of the biological importance of extrinsic processes as motivation for the development of our approach, the first well-established, and the second considered here. First, although intrinsic noise is an important contributor, extrinsic processes are known to be a substantial and sometimes dominant
source of variation in gene expression levels across cells and over time \cite{Elowitz2002,Raser:2004gh,Rosenfeld:2005hn,Zechner:2014}. We are now beginning to understand the
underlying biological sources \cite{DasNeves:2010km}, which
include effects related to circadian oscillations, temperature, chromatin remodelling,
the cell-cycle and pulsatile transcription factors \cite{Eser:2014,Levine:2013}. To understand gene expression, it is therefore essential to move beyond the SSA, which can only account for intrinsic noise, and to include other sources of variation. Second, fluctuations in the expression, degradation and recycling of proteins inevitably affect the way networks containing those proteins function and the extent of stochasticity in the input they provide to other networks. Fluctuations in the component proteins of signal transduction networks limit information transfer \cite{Cheong:2011jp}, affect transduction network `design' \cite{Bowsher2014} and, although often overlooked, are inevitably conveyed (as extrinsic inputs) to the networks regulated by signaling. Here, the computational advantages of Extrande will allow us to demonstrate how fluctuations in the protein componentry of signal transduction networks are conveyed to signaling outputs and place strong constraints on the design of networks determining cell fate, thus influencing the distribution of phenotypes at the population level. Without the ability to simulate biomolecular networks that are exposed to fluctuating inputs, the ability to address such questions is severely restricted.


{\bk There are two existing approaches to stochastic simulation of reaction networks subject to dynamic, fluctuating inputs.  
The first class of algorithms \cite{Gillespie:1976,Rand:2012,Guerriero:2012} implements the SSA, under the \emph{approximation} that the input remains constant between the occurrences of any two reactions. However, this approximation can give spurious results even when dynamic inputs to the network are changing
relatively slowly. We term these collectively the {\bk \emph{Slow Input Approximation method} (SIA).} 
The second class of algorithms \cite{Anderson:2007,Shahrezaei:2008iq,Johnston:2012} involves step-wise numerical integration of reaction propensities until a target value for the integral is reached. {\bk Algorithms in this class would be (conditionally) exact, if it were not for the presence of numerical error in integration, but can impose large and impractical computational burdens, especially when cell ensembles are studied}. We term these collectively the \emph{integral method} (distinguishing next and direct integral approaches below). We perform a comparative analysis of both methods with Extrande and demonstrate that our method offers an accurate and computationally efficient alternative approach. Extrande involves no analytical or numerical integration but instead relies on `thinning' techniques \cite{lewis1979,OGATA:1981}. {\bk Other approaches using rejection methods have also recently been proposed as a means to tackle systems with time-dependent propensities \cite{Zechner2014,Thanh2015}.}



\section*{Results}

\subsection*{Stochastic simulation using the Extrande approach}

{\bk The stochastic simulation algorithm (SSA) \cite{Gillespie:1976,Gillespie:1977ww} allows simulation of biomolecular reaction networks taking into account the discreteness of these systems as well as the intrinsic randomness in the timing of reaction events. The SSA assumes that the propensity of each reaction channel to fire,  hence the probability of the reaction to occur over a small time interval, remains constant between reaction events. This naturally restrains the use of SSA to simulate networks embedded in dynamic, fluctuating environments because the reaction propensities then become time-varying quantities under the influence of extrinsic processes.}

{\bk \begin{center}
\doublebox{\begin{minipage}[b]{0.95\columnwidth}
\subsection*{Box 1:  Extrande algorithm  }
\small{
We present below the Extrande algorithm---\textbf{\emph{Ext}ra Reaction Algorithm for
Networks in Dynamic Environments}---for stochastic simulation
over the interval $[0,T]$ of a reaction network with $M$ reaction
channels $\{R_{1},...,R_{M}\}$ and associated stoichiometries $\{v_{1},...,v_{M}\}$.
The network state, $X$, gives the number of molecules of
each species. We denote the extra (`virtual')
reaction channel by $R_{M+1}$. The algorithm takes as input a function
that simulates the dynamic, exogenous inputs, $I$, over time (see
below). The variable $t$ below tracks the progress of the algorithm in continuous time. 
\vspace{0.05in}
\begin{algorithmic}[1]
\State Initialise time $t \gets 0$ and network state $X \gets X_0 $.
\Repeat 
\State \parbox[t]{0.95\linewidth}{({\it Determine propensity bound}) Choose $ L\leq T- t$ and $B$ such that $a_{0}(t+u)\leq B$ for $0\leq u<L$, where $a_{0}(t+u)=\sum_{j=1}^{M}a_{j}[X,I(t+u)]$ is the sum of the reaction propensities $a_{j}$ at time $t+u$ provided that no reaction channel fires  during $(t,t+L)$.}
\State ({\it Generate putative reaction time}) Draw exponentially distributed random number $\tau \sim \mbox{Exp}(1/B)$.
	\If{$\tau > L $}    
	\renewcommand{\Statex}{\item[\hphantom{Step.}\hphantom{Step}] }
		\Statex{({\it`Reject'; State of the network remains unchanged})}
		 \State{Update time $t \gets t + L $. }
	\Else{}
		\State{Update time $t \gets t + \tau $.} 
		\State{\parbox[t]{0.9\linewidth}{From the simulation of $I$ at time $t$, obtain $I(t)$, update all  propensities $a_{j}[X,I(s)]$ that depend on $I(s)$, and evaluate the sum $a_{0}(t) = \sum_{j=1}^{M}a_{j}[X,I(t)]$.}}
		\State{Generate uniformly distributed random number $u~\sim U_{(0,1)}$.}
		\If{$a_{0}(t)\geq Bu$ } 
		\renewcommand{\Statex}{\item[\hphantom{Step}\hphantom{Step}\hphantom{Step}] }
		        \Statex{({\it`Accept'; Choose reaction channel to fire and update state})}
			\State{choose reaction associated with the smallest positive integer $j$ less than or equal to $M$ satisfying:}
			\vspace{-0.1in}\[\sum_{i=1}^{j}a_{i}[X,I(t)]\geq Bu,\]\vspace{-0.1in}
			\State Update state $X\leftarrow X+\nu_{j}$.
		
		\Else{} 
		\Statex ({\it`Thin'; The extra reaction channel fires and the state of the network remains unchanged})
		\EndIf
	\EndIf
\Until{$t \ge T$} ({\it terminate when final time is exceeded} )
\end{algorithmic}
\vspace{0.05in}}
The function used to simulate the inputs, $I$, will depend on the input processes. For example, when $I$ is given by a stochastic or ordinary differential equation (SDE or ODE) requiring numerical solution, the function returns values of $I$ on a discrete grid (using, e.g., the Euler-Maruyama method \cite{higham2001} in the case of an SDE), with values for intermediate times obtained by a deterministic interpolation rule. Notice that, in general, the bound, $B$, and look-ahead horizon, $L$, change on each repeat of the algorithm: both may depend on the history of $X$ at time $t$ and on the trajectory of $I$ on $[0,T]$. The bound, $B$, may be set to the supremum of $a_0(t+u)$ for $0 \le u<L$:  e.g., in the case of single input and with all $a_j$ monotonically increasing functions of $I$ this would be $B=\sum_{j=1}^{M}a_{j}[X(t),I^*]$ where $I^*$ is the supremum of $I(t+u)$ for $0 \le u<L$. Different methods for computing the bound $B$ and various implementations of Extrande are given in SI. A simple choice for the look-ahead horizon is $L=T-t$. In practice, we find that the efficiency of the method is relatively robust to the choice of $L$ and a few exploratory simulation runs can guide its choice (see Fig. 2). Alternatively $L$ could be adaptively updated at the beginning of each repeat based on information collected by the algorithm (e.g.,  statistics of `thin', `reject' and `accept' events).
\end{minipage}}
\end{center}
}

{\bk Extrande (Box 1)---or \emph{Ext}ra Reaction Algorithm for Networks in
Dynamic Environments---allows exact stochastic simulation of any downstream
reaction network, conditional upon a time course of the dynamic inputs that is simulated up-front. The method involves no analytical or numerical integration, though we give a connection to the direct integral method below, and instead makes use of point process `thinning' techniques \cite{lewis1979,OGATA:1981}, where some simulated events are discarded. The only error incurred is any error associated with the input pre-simulation, typically an approximate simulation of a stochastic differential equation (Box 1). 

The Extrande approach can be understood as introducing an extra, `virtual' reaction channel into the system (whose occurrence
does not change molecule numbers). The propensity of the extra channel is designed to fluctuate over time so that (when added to the sum of all other reaction propensities) the total propensity in the \emph{augmented} system becomes constant between events and equal to an upper bound on the sum of the propensities in the \emph{original} system. To accomplish this, the method exploits the exogeneity of the dynamic inputs---the assumption of negligible retroactivity \cite{DelVecchio:2008gy} from network to inputs.
In particular, their exogeneity means that Extrande is able
to make use of the `future' trajectory of the inputs to find an upper bound, $B$, on the total propensity, which is valid over a certain time interval $L$ (see {\bk Step 3}, Box 1). 

Simulation of the \emph{augmented} system is feasible by means of an SSA-like algorithm. The method uses the bound on the total propensity to generate a putative reaction reaction time $\tau$ (Step 4). If the reaction time exceeds the time horizon $L$, it is rejected; the system time advances by $L$ (Step 6), and the procedure restarts by determining a new bound. Otherwise, time advances by $\tau$ and a reaction is chosen based on the updated reaction propensities (at time $t+\tau$) (Steps 8-15). The reaction events of the virtual channel are discarded, leaving those of the other channels---because the simulated timing and types of the biochemical reaction channels are unaffected by the behaviour of the extra channel, the result is a trajectory of the original system (see Methods).}


\subsubsection*{The Extrande method is accurate but the {\bk SIA}  method can fail, even when inputs vary relatively slowly}

Under the {\bk SIA}  method (see SI), the input is approximated by a piecewise constant
function whose value can only change when the firing of a biomolecular reaction is simulated. 
The method does not track the instantaneous value of the input process but values of its past; the process simulated therefore becomes non-Markovian.
Nevertheless, one might expect that the {\bk SIA}  method would be adequate when the input changes on
a slow timescale, compared to the typical waiting times between
firings of the physical reaction network when exposed to the input \cite{Gillespie:1976}. We demonstrate (Fig. 1) this is far
from being the case using gene expression models with dynamic transcription propensities, and biologically realistic protein abundances
and rates for various cell types: unicellular algae, bacteria, yeast, and mammalian cells. 
{\bk Specifically, we consider the two-stage model   
\begin{align*} \varnothing \xrightarrow{k(t)} M, \ \ M \xrightarrow{k_s} M + P, \ \ M \xrightarrow{k_\mathrm{dm}} \varnothing, \ \ P \xrightarrow{k_\mathrm{dp}} \varnothing, 
\end{align*} with time-varying transcription propensity, $k(t)$. The translation rate, $k_{s}$, and the mRNA and protein degradation rates, $k_\text{dm}$ and $k_\text{dp}$ respectively, are constant parameters.}

We focus on two important timescales for changes in transcription rates, the circadian 24 hour period \cite{panda2002} and the length of
the cell cycle \cite{Eser:2014}. For the unicellular alga \emph{O. tauri}, a model organism for circadian
rhythms \cite{Millar:2011}, the error made by the {\bk SIA}  method (in predicting average expression
by a cell population) when the transcription rate follows the circadian
sinusoid is conspicuous ($>60\%$) across the entire physiological
range of protein abundances---despite there being just 0.008
circadian cycles per protein lifetime (and 0.002 cycles per mRNA lifetime) for this organism (Fig.~1A).
{\bk For processes with stationary, fluctuating transcription rate, the error depends on the correlation time, $\gamma$, (Fig.~1B) which is of the order of cell cycle. Typical parameters in bacteria yield particularly large ($60-90\%$) errors in the mean expression, again across the entire physiological abundance range.} 
We have verified, throughout Fig.~1 A, C \& D, close agreement of the results generated using Extrande with the 
corresponding analytical results (derived as in \cite{Bowsher2013}; SI, Fig.~6). All simulations were performed (for the Extrande, integral and {\bk SIA}  methods) using a modified version of the iNA software \cite{Thomas:2012}. 
{\color{black} We provide an implementation of Extrande and the {\bk SIA}  method reproducing the results of Fig. 1C (SI S1).}

\begin{figure*}[h!]
\vspace*{-.5in}
\begin{center}
 \includegraphics[width=0.80\textwidth]{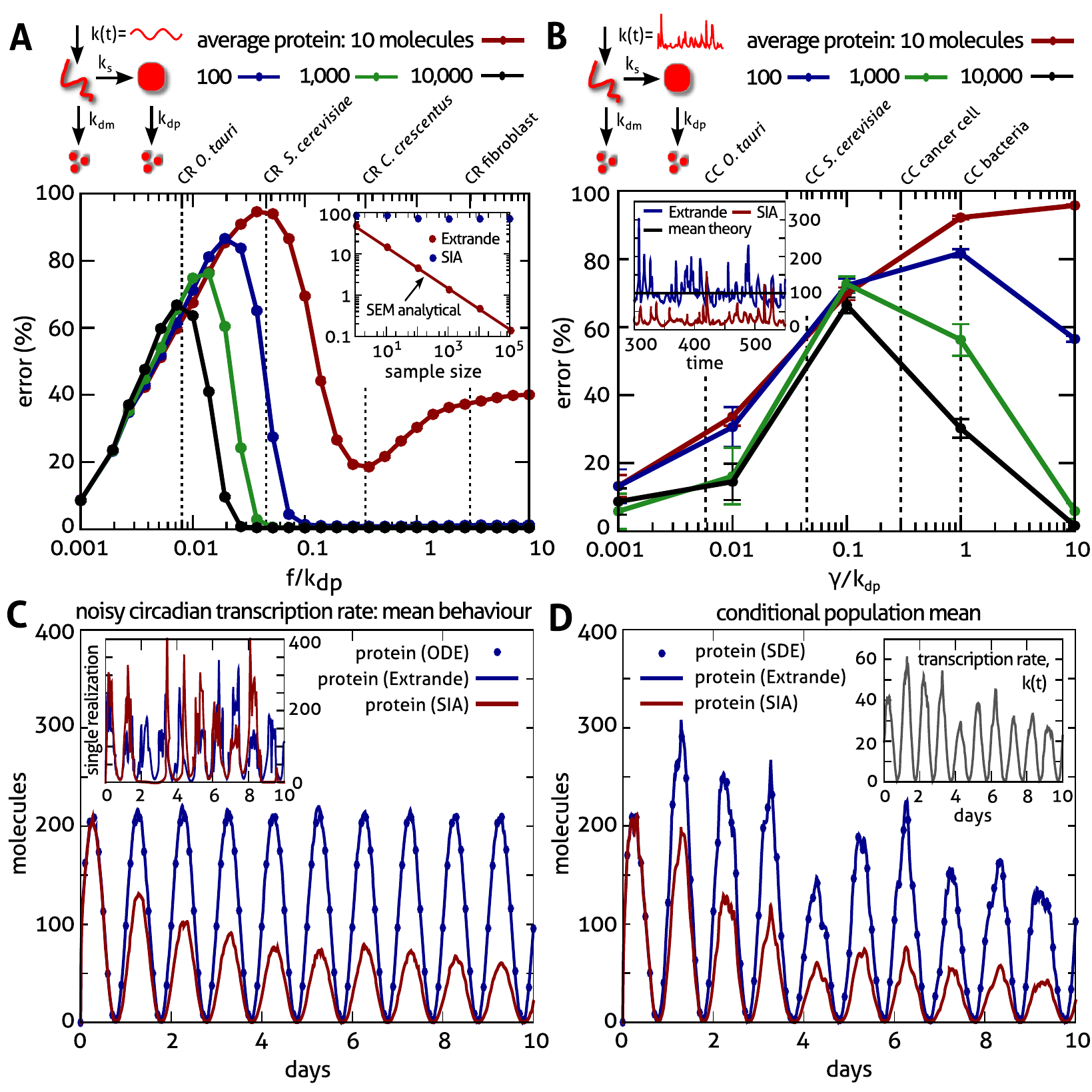} 
\caption{ \small
 \textbf{Comparison of the accuracy of the Extrande and {\bk SIA}  simulation methods}. {\bf{(A)}} \textbf{Gene expression with circadian transcription rate}, {\color{black} $k(t)/k_{dp}=4(1+\sin(2\pi f t))$}, and period $f^{-1} = 24$h. The proportional root mean square error between the average protein number from the {\bk SIA}  method, $\langle n(t) \rangle$, and the exact, time-dependent solution, $n_\text{ex}(t)$, is shown as a function of the relative frequency of oscillation. {\bk The error is given by $\left[ \frac{1}{T \bar{n}_\text{ex}^2} \int_0^T \mathrm{d} t \, (\langle n(t) \rangle-n_\text{ex}(t))^2\right]^{1/2}$, where $\bar{n}_\text{ex}$ is the time-average of the exact solution.} Physiological parameters for circadian rhythms (CR) in 4 different cell types are indicated. Actual mean protein numbers are varied via the translation rate ($k_s$), holding degradation rates constant. The error is particularly conspicuous ($>60\%$) for \emph{O. tauri} over the whole range of average protein numbers ($10-10,000$) whereas the exact Extrande method (Inset) accurately predicts the mean protein numbers for this case within sampling error (given by the standard error of the mean, SEM). {\bf{(B)}} \textbf{Gene expression with noisy transcription rate}, $k(t)=\langle \exp(\xi(t) \rangle^{-1} \exp(\xi(t))$ where $\xi(t)$ is zero-mean Gaussian (OU) noise with autocovariance $\langle\xi(t)\xi(t')\rangle=5 e^{-\gamma|t-t'|}$. We show the proportional error of the stationary average protein number from the {\bk SIA}  method as $\gamma$ varies, with average protein numbers set via $k_s$ as in (A). {\bk Autocorrelation times} of the transcription rate of the order of the cell cycle (CC) are indicated for 4 different cell types. The error is particularly conspicuous ($60-90\%$) for stable proteins removed mainly by dilution, as is common in bacteria ($\gamma/k_\text{dp}=1$), where we show (Inset) simulated average protein numbers for a population of 100 cells. {\bk The error bars denote one standard deviation of the bootstrap distribution.} {\bf{(C)}} \textbf{Noisy circadian oscillations in an \emph{O. tauri} cell population}. Average protein numbers for $2500$ cells and $10$ days simulated using a circadian transcription rate with cell cycle-induced amplitude fluctuations on a similar timescale: $k(t)=20\exp(\xi(t))(1+\sin(2\pi f t))$, where $\xi(t)$ is zero-mean Gaussian (OU) noise with autocovariance  {\color{black} $\langle\xi(t)\xi(t')\rangle=\frac{1}{8}e^{-\gamma|t-t'|}$ and $\gamma/k_\text{dp}= f \ln2 $}. While Extrande correctly predicts sustained oscillations (blue), the {\bk SIA}  method predicts only damped oscillations (red). Extrande is in excellent agreement with the corresponding moment equations of the master equation (dots, equivalent to ODE solution). Single cell realizations (Inset) reveal the {\bk SIA}  method shows unphysical loss and revival of oscillations. {\bf{(D)}} {\bk \textbf{
Average behaviour of \emph{O. tauri} cells conditional on transcription dynamics:
 }}  We pregenerated a single realization (Inset) of the transcription rate, $k(t)$, used in (C), and averaged over $1,000$ resultant protein trajectories (all parameters as in C). The solution of the corresponding SDE for average protein conditional on the trajectory of $k(t)$ agrees very well with the average from Extrande, in contrast to the {\bk SIA}  method. See SI for simulation details and other rate parameters.}
\end{center}
\end{figure*}

{\bk The error of the SIA method in Fig.~1~A \& B depends non-monotonically on the input frequency. While small errors are expected for extremely slow inputs, the method performs well also for comparably fast inputs and large molecule numbers because the system effectively averages the signal. In the intermediate regime, the error is considerable and the SIA method yields qualitatively misleading results (Fig.~1~C), predicting damped rather than sustained oscillations of the average protein expression.} The damping arises from
loss of protein expression in individual cells (inset Fig. 1C), which is
not always reinstated. We find that {\bk SIA}  error plots in Fig. 1A are well explained by the difference between the fractions of time during which protein and mRNA numbers are both zero in the {\bk SIA}  and Extrande simulated trajectories respectively (SI, Fig.~5). When protein and mRNA numbers are zero or near-zero, the transcription rate fails to update or sluggishly updates under the {\bk SIA}  method, since the value of its input only updates when some reaction
channel fires. 
In the extreme case of zero copy numbers
occurring while the true transcription rate is zero, the {\bk SIA}  method gets
trapped and simulates no further reactions.
{\bk We presume this reasoning also explains the misleading damping predicted by SIA for the average protein concentration conditional on a particular trajectory of the transcription rate (Fig. 1D), whereas in reality the conditional mean closely follows the input dynamics due to the linearity of the system and its fast dynamics (compared to the circadian timescale).}  

A {\bk SIA}  algorithm could be considered in which the input is updated on either a predetermined or random grid, with the grid resolution chosen in advance on the basis of the time-scales in the network. Such an algorithm is expected to be computationally demanding for {\bk systems in which stiffness arises due to rapidly varying inputs---the grid must be fine to account for this timescale}, while the simulation time will have to be long to account for the largest timescale, resulting in a very slow algorithm. {\bk By contrast, the performance of the Extrande algorithm is limited only by the firing times of the extra reactions and hence by the quality of the upper bound. It thereby avoids \emph{ad hoc} discretization schemes and the need to experiment with multiple choices of resolution.}

\subsubsection*{The Extrande method can speed up simulation by several orders of magnitude compared to the integral method  }

The Modified Next (MN) integral method has been proposed \cite{Anderson:2007} as well-suited to simulation when there are
time-varying propensities. We obtained a breakdown (Fig.~2A) of the CPU time
of Extrande and compared this to the CPU time of the MN integral method (with the same time-step used for integration and for up-front simulation of the input). We use the noisy circadian transcriptional input and network in Fig. 1D as an example.

\begin{figure*}[h!]
\vspace*{-.1in}
\begin{center}
 \includegraphics[width=1\textwidth]{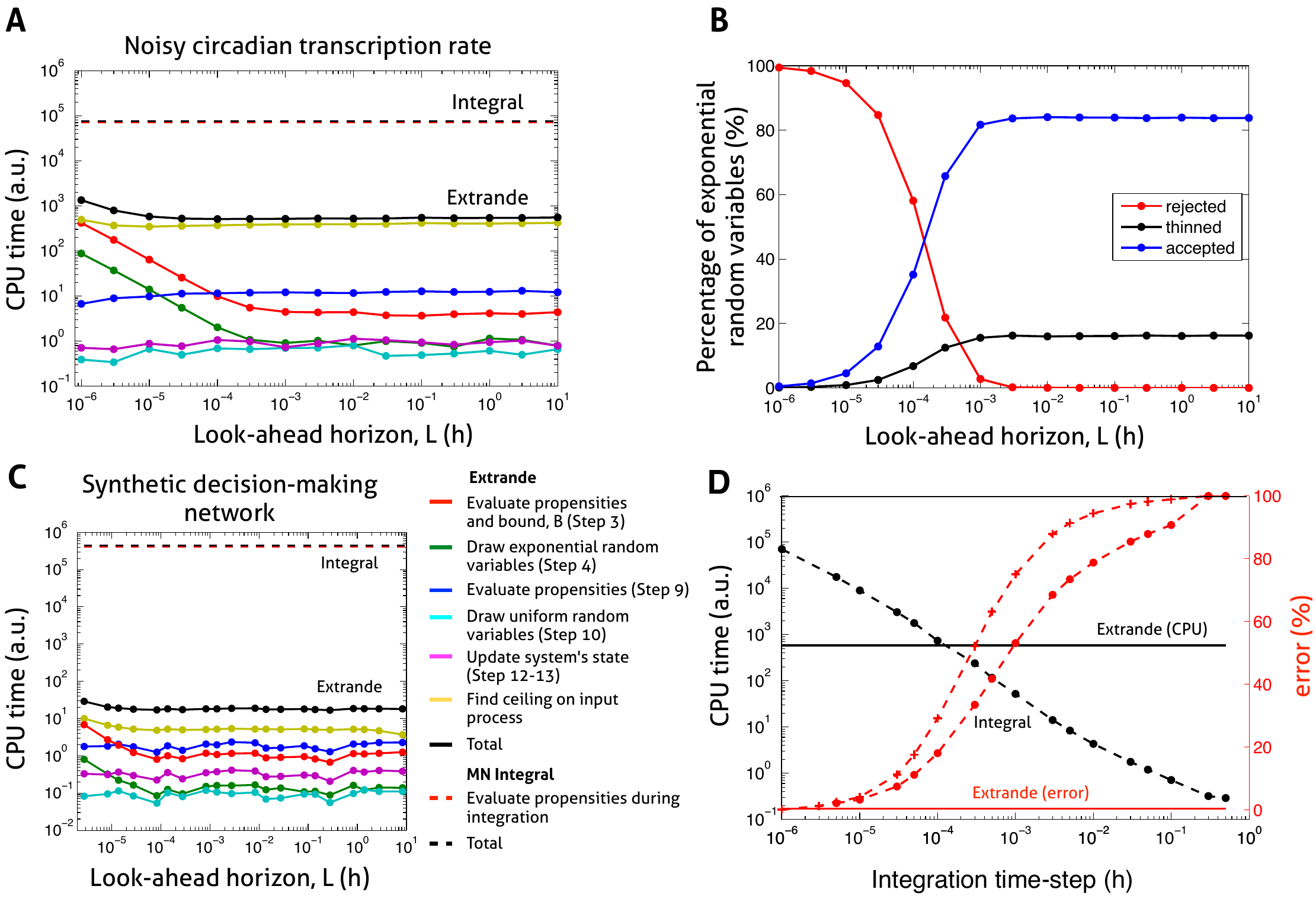}
\caption{ \small
\textbf{Comparison of the Extrande and integral methods}.
{\bf{(A)}} Comparison of CPU times for Extrande and the modified next (MN) integral method \cite{Anderson:2007}. CPU times broken down into their constituents (color coded), and shown as a function of the look-ahead horizon, $L$, for Extrande (see also Box 1). Time-step of input presimulation and of integration for the MN method both equal to $10^{-6}$h. 
{\bk CPU times were collected while simulating the two state model of gene expression with noisy circadian transcription (see Fig. 1D) up to $t=10$ days.} 
{\bf{(B)}} Percentage of exponential random variables generated in {\bk Step 4} of Extrande (Box 1) that are rejected, thinned, and accepted, as a function of $L$. Extrande simulation, network and input as in (A).
{\bf{(C)}} As in (A) but for the SynDM network (Fig. 3A) with single OU input presimulated using a time-step of $10^{-2}$s (and with lifetime 1h, CV 0.5), and integration time-step for the MN method also equal to $10^{-2}$s.
{\bf{(D)}} Comparison of CPU times (for 10 simulated days) and of percentage errors for Extrande and the MN integral method. Network and input as in Fig.~1D (and panels A \& B), time-step of input presimulation again equal to $10^{-6}$h. The absolute value of the percentage error in the integral method's estimate of the conditional mean is shown in red, both at 6h (crosses) and averaged over the first 24h (compared to Extrande, circles). CPU time for Extrande corresponds to an intermediate value of $L$ (in practice, a few of the 1000 cells would be run initially to choose $L$). Throughout Fig.~2, we use trapezoidal numerical integration for the MN integral method; {\bk the implementation of Extrande uses input presimulation over the look-ahead horizon $L$ from which its ceiling value is obtained}; and the CPU time for input presimulation is excluded since it is identical for the MN and Extrande methods.}
\end{center}
\end{figure*}

The MN integral method has a CPU time 140 times that of Extrande (with intermediate look-ahead horizon, $L$)---4.6 months, for example, is reduced by Extrande to 1 day (Fig. 2A). The source of the improvement is the substantial reduction by Extrande in the CPU time spent on propensity evaluation, which accounts for the vast majority of the total CPU time of the MN integral method. Breakdown of the total CPU
time of Extrande reveals
that it is dominated by the time spent finding a local ceiling on the input
trajectory. This computational cost, however, is more than outweighed by the reduced CPU time spent on propensity evaluation. The total CPU time of Extrande is mostly insensitive to $L$, the fixed look-ahead horizon used (except at smaller values of $L$). Recall that {\bk Step 4} of the Extrande algorithm (Box 1) generates exponential random variables (waiting times). Smaller values of $L$
are associated (Fig.~2 A\&B) with a higher proportion of rejected exponentials, a lower proportion of `thinned' exponentials (those resulting in firing of the extra channel), and higher
CPU times incurred in evaluating propensities and drawing exponential
random variables (in {\bk Steps 3 and 4} of the algorithm). We observe similar behaviour of the CPU times and CPU components of Extrande (Fig.~2C \& SI, Fig.~7) for simulation of the synthetic decision-making network studied below---using equal integration and input presimulation time-steps, the MN integral method has a CPU time 25,000 times that of Extrande (with intermediate look-ahead horizon, $L$), with the vast majority of the CPU time for the integral method again spent on propensity evaluation.

We also compared the CPU times and accuracies of the MN integral method to those of Extrande for a range of time-steps of numerical integration (Fig. 2D), again using the input and network of Fig. 1D. In practice, of course, multiple integration time-steps would require investigation to assess convergence (and there would usually be no analytical result available with which to make
comparison). For time-steps giving an absolute relative error
$< 5$ \%, the CPU time for the MN integral method is at least 15 times the
CPU time for Extrande, which has the lower relative error. For
integration time-steps resulting in equal CPU time for Extrande and the MN integral
method, the relative error of the latter is 30\% (at the 6h point). We note that the time-step used to presimulate the noisy transcriptional input ($10^{-6}$h) is sufficiently small to ensure an error near $0$\% for Extrande. We also show (see SI), again in the context of Fig. 2A, that the CPU time of the direct integral method is also expected to exceed the CPU time of Extrande. For the MN integral method, an integration time-step equal to that for input presimulation can in theory be used to again leave only the error associated with input simulation but such integration time-steps can make simulation of the model computationally infeasible (SI, Fig.~7B).

It is clear that the Extrande method offers important advantages compared to integral methods in terms of simulation speed. Furthermore, Extrande avoids the need to assess convergence of estimates as the time-step of integration in decreased. The total reduction in CPU time can be enough to make a previously infeasible simulation project computationally practical. We present results for such a project below (Fig.~3) that consumed 2.3 months of computing time using Extrande but we calculate would have taken in excess of 14 years computing time using the integral method. The goal was to simulate the distribution between 2 phenotypes in a population of 1000 bacterial cells responding to stress conditions (at the end of a 20 hour experiment in calendar time). The `competence' networks of interest decide cell fate in a stochastic fashion and have attracted considerable attention, not least as a model of differentiation. However, these networks are regulated by upstream quorum signaling and this regulation has not been studied quantitatively---it turns out to be essential for understanding the wild-type design, not only to model the networks stochastically, but also to allow for fluctuations from the upstream signaling.

\subsection*{Robustness to extrinsic fluctuations from upstream signaling constrains the design of cell fate networks}
We study the decision to enter competence (for uptake of extracellular
DNA) by the model organism\textit{ Bacillus subtilis.} It is well established \cite{Maamar:2007,Suel:2006ea,Suel:2007gp} that the source of differentiation of 10-20\% of the cell population
under stress conditions is fluctuations in transcription of the master competence regulator, ComK.
The ComS-MecA-ComK competence module is regulated by the
activated transcription factor pComA, the output of the transduction
mechanism relaying extracellular, quorum sensing signals (CSF and
ComX), see Fig.~3A. We study the effect of this upstream signaling on differentiation into the competent phenotype.

A useful approach to understanding the structure-function relationship in systems biology is to rewire networks found in nature and compare function with the wild-type, which can then shed light on why apparently similar network structures were not adopted naturally \cite{Cagatay:2009ie}. 
In the wild-type, upstream signaling acts via activation of the \textit{ComS promoter} by pComA binding (Fig.~3A, thick black arrow). We compare the behaviour of wild-type cells to those with a Synthetic Decision-Making network (SynDM) which is regulated, in addition, via activation of the \textit{ComK} \textit{promoter}
by pComA binding (red dashed arrow). {\bk We model ComK-driven progress and entry into functional competence, and write 
$\text{Progress}(t) = k \int_0^{t} \mathrm{ComK}(s) ds$,
where $k$ is an effective rate of ComK-driven differentiation. A cell is taken to enter (functional) competence at the time when $\mathrm{Progress}(t)=1$.  The value of the parameter $k$ is set so that the wild-type and SynDM networks give equal fractions of competent cells with a constant level of pComA (1000 molecules).} 
We tune rate parameters associated with the ComK promoter of the SynDM network so that the fraction of SynDM cells entering competence (0.18)
is the same as for wild-type cells, in the absence of
fluctuations in pComA levels (see SI). {\bk A table listing all reactions and parameter values used in our models of the competence module of wild-type {\it B. subtilis} and the SynDM networks is given in the SI}.

We use the linear noise approximation (LNA) \cite{Kampen:1961} to model the the upstream signaling (with CSF and ComX fixed at steady-state levels), giving a mean for pComA of 1000 molecules throughout. Importantly, we include in the model gene expression and degradation of the proteins comprising the signal transduction mechanism because it is now understood that the resultant variation has important effects on signaling and information transfer{\bk~\cite{Bowsher2014}}. A single Ornstein-Uhlenbeck (OU) process is sufficient to closely match the mean, variance and autocorrelation function of pComA given by the LNA (see SI). We therefore use a single OU process for the pComA input in what follows. A range of protein lifetimes is considered, consistent with the broad range of cell-cycle periods observed for bacteria under
different growth conditions \cite{Michelsen:2003}, where nutrient
limitation can result in periods in excess of 10h. Our baseline LNA model of the upstream signaling module gives a lifetime and CV of pComA fluctuations equal to 5h and 0.35. We take the pComA input to be exogenous to the ComS-MecA-ComK competence module since it is in high abundance relative to the 2 promoters it binds (the only interaction between the two modules).

The importance in determining cell fate of the time taken for the cell to complete different differentiation programs (to the point of irreversible commitment) has recently been emphasised \cite{Kuchina:2011}. The SynDM network creates a differentiated sub-population by activating
the differentiation program in most or all of the cell population (Fig.~3 C\&D), with entry to competence the outcome of a `race' to differentiate
over the relevant time window.  In the SynDM network, binding of pComA
to the ComK promoter results more often in periods of non-zero
ComK expression than in the wild-type population, but when such periods occur, they are less sustained
(Fig. 3B-D \& SI, Fig. 8). The typical rate of progress of a SynDM cell to competence
is increased by a higher level of pComA (SI, Fig.~8), and extrinsic fluctuations in the pComA level therefore affect the fraction of cells entering competence (Fig.~3 C\&D). In
contrast, the wild-type activates the differentiation program in a
smaller sub-population, the size of which is under modest regulation by pComA (Fig.~3F)---a high proportion of the active wild-type cells then enter competence because, once activated,
ComK expression rarely deactivates in the wild-type (Fig.~3B \& SI, Fig.~8).

We find two important advantages of the wild-type design (in addition
to the implied reduction in the metabolic cost of gene expression). First, the fraction of
cells entering competence is considerably more robust to the 
fluctuations from upstream signaling in pComA (Fig.~3E). For example, with the baseline model of upstream signaling,
the SynDM network has a competent fraction (40\%) which is more than 2.25 times the
competent fraction when pComA is held constant at its mean level, whereas the competent fraction of wild-type cells (17\% cf 18\%) has changed very little.
The difference in robustness is explained by the sensitivity of the probability of competence
for a SynDM cell as a function of the time average of the signal, $\langle \mathrm{pComA}\rangle$,
which switches quite rapidly from zero to one (Fig.~3F). Since the
fraction of competent cells is equal to the average of $\mathrm{Prob}[\mathrm{{Competence}}| \langle \mathrm{pComA}\rangle]$
over the distribution of $\langle \mathrm{pComA}\rangle$ (which
is approximately the distribution of pComA for longer lifetimes), the
competent fraction \emph{increases} in the presence of extrinsic fluctuations
for SynDM (recall the mean of pComA is 1000 molecules). In contrast, $\mathrm{Prob}[\mathrm{{Competence}}|\langle\mathrm{pComA}\rangle]$
is approximately linear for the wild-type network, which implies that
the competent fraction depends largely on the mean of pComA alone. Such
plots (Fig.~3F) should prove a useful diagnostic tool for the design of synthetic decision-making
networks.

The second advantage of the wild-type design is that the
fraction of cells entering competence is also considerably more robust
than SynDM to heterogeneity across the cell population in the rate at which
ComK-driven differentation proceeds (Fig.~3G). The reason is evident from the progress to competence trajectories in Fig. 3B-D. We note that fluctuations from upstream signaling in pComA can also cause decreases in the
fraction of competent SynDM cells, as seen for higher rates of differentiation (Fig. 3G). Heterogeneity in the rate at which differentiation programs proceed
is inevitable where cellular decisions are executed by large gene expression
networks and involve substantial physiological changes \cite{Hahn:2005}.

These \textit{in silico} experiments (Fig.~3), made computationally feasible by Extrande, cast light on the wild-type network design in which quorum signaling input to the competence decision-making network (ComS-MecA-ComK) by the transcription factor pComA exerts its effect only at the promoter of ComS and not at the promoter of ComK. The experiments reveal exquisite robustness of the wild-type design to fluctuations from upstream signaling and to heterogeneity in downstream processes, and 
demonstrate the computational potential of Extrande for \emph{in silico} network design.  

\section*{Discussion}

Stochastic simulation of biomolecular networks is now indispensable for studying
biological systems, from small reaction networks to large ensembles of cells. The effects of stochasticity can be pervasive at the single-cell level, determining the distribution of phenotypes in a population and thus potentially affecting evolutionary outcomes. However, studying such phenomena requires stochastic simulation of a large ensemble of cells that can take into account both intrinsic and extrinsic sources of cellular variation. This can be hugely costly in terms of CPU time, placing important \emph{in silico} experiments out of reach. Here we provide the new Extrande approach---for stochastic simulation of a biomolecular network embedded in 
the dynamic environment of the cell and its surroundings---which substantially increases the computational feasibility of such experiments without compromising accuracy.

We show that previous approaches to this problem either can fail dramatically, even when inputs vary relatively slowly, or impose impractical computational burdens due to costly numerical integration of reaction propensities. 
Given a simulated trajectory of fluctuating network inputs, the Extrande approach provides a conditionally exact
solution that can speed up simulation by several orders of magnitude compared to integral methods. 
{\bk In practice, we find that integral methods suffer from the high cost of propensity evaluations during numerical integration. Extrande bypasses numerical integration by introducing an extra reaction channel---one designed to keep the total propensity of the `augmented' system constant between events---hence making the problem of finding the time to the next event analytically tractable.
Importantly, our numerical results demonstrate that the overhead costs induced by the Extrande method---for example, due to 
thinning 
and rejection 
events, and due to obtaining the ceiling of the input process when a global ceiling is not available--are significantly lower than the cost of accurate numerical integration. In practice,} we observe speed-ups by a factor as great as $2.5 \times 10^{4}$ {\bk (Fig.~2C)}.

Recent work \cite{Zechner2014} proposes to handle fluctuating environments in a different manner, by deriving a network model for the biochemistry that takes account of the dynamic input and follows the correct (marginal) probability law. Explicit simulation of the input is bypassed. The resultant `uncoupled' network model has time-varying reaction propensities and can then be simulated using integral or thinning methods. However, analytical derivation of the uncoupled network model is not always possible, particularly when there are multiple inputs. The accuracy of the method then depends on finding suitable approximation schemes.


We exploit the benefits of the proposed Extrande simulation method here to study the decision-making behaviour of a quorum sensing population of bacterial cells. {\bk The \textit{in silico} experiments presented (Fig.~3) took approximately two computing months using Extrande (and an Intel Xeon, 3.3GHz quad-core processor with 32GB of RAM), but would have been prohibitive using the integral method due to the approximate $70$-fold slow down needed to ensure even modest accuracy \bk (see SI, Fig.~7B)}. The results elucidate the costs and benefits of alternative network designs for the probabilistic differentiation of a sub-population of cells in response to upstream signaling. Our findings argue for the biological significance of fluctuations in signaling inputs that arise from synthesis and degradation of the protein componentry of signal transduction networks, and show that these fluctuations have important consequences for downstream networks such as those deciding cell fate. 
We expect the accuracy and reductions in CPU time made possible by Extrande to help open up the landscape of computationally feasible simulation of biomolecular networks and cell ensembles. Extrande thus has the potential to accelerate both understanding of molecular systems biology and the design of synthetic networks.

\begin{figure*}[h!]
\vspace*{-.3in}
\begin{center}
\includegraphics[width=0.75\textwidth]{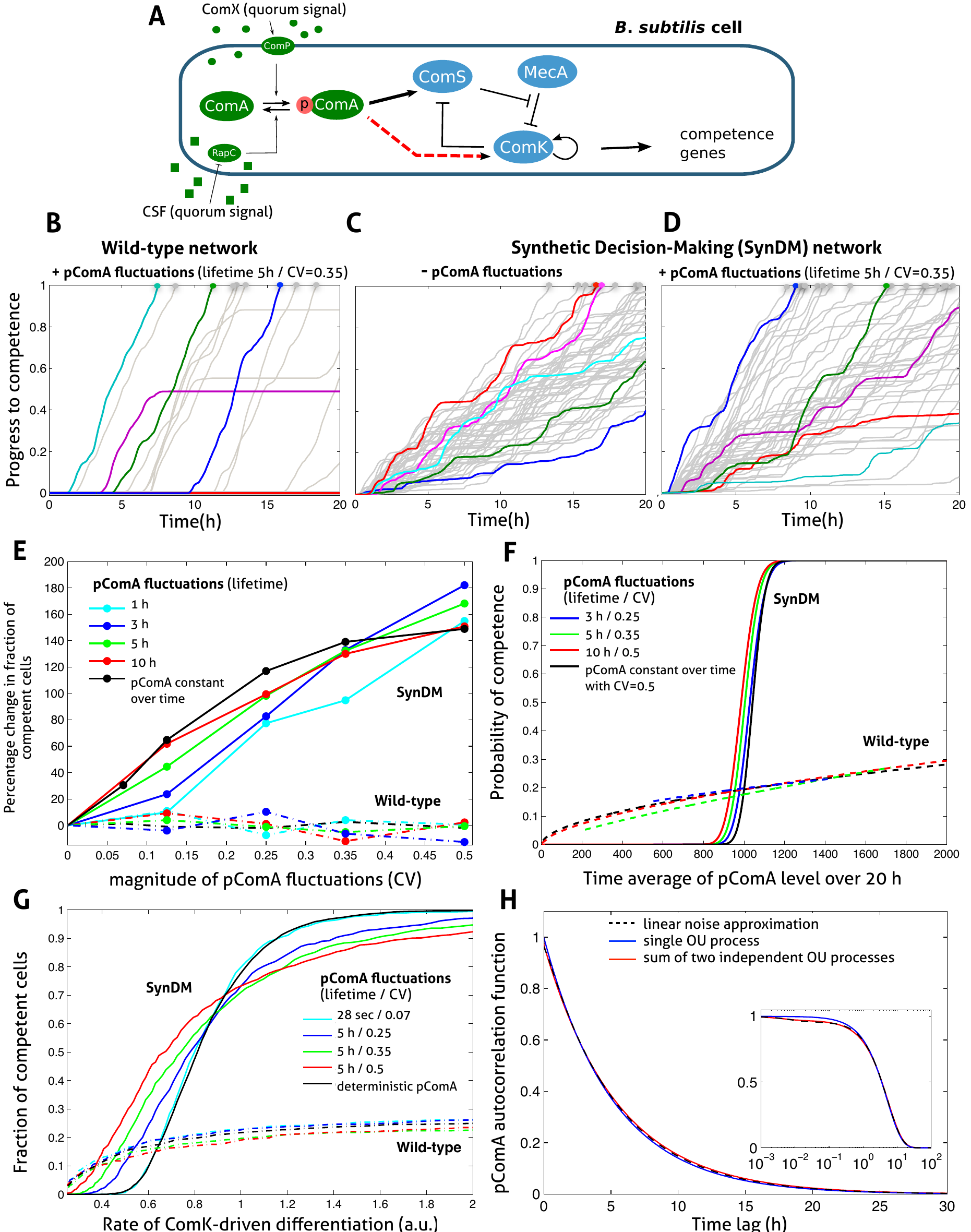}
\vspace*{-.1in}
\caption{\small \textbf{The effect of extrinsic fluctuations from upstream quorum
signaling on the competence decision of \emph{B. subtilis}}.
{\bf{(A)}} The wild-type signaling
(green) and competence (blue) modules. The Synthetic
Decision-Making network (SynDM) has the additional positive regulation
of ComK by pComA (dashed red arrow). Reaction networks and rate parameters described
in detail in the SI. {\bf{(B--D)}} Time courses of progress to
competence shown for 100 cells containing the wild-type and SynDM networks,
simulated using Extrande. In B \& D, independent Gaussian, OU input processes for
the pComA level in each cell are used, derived from an LNA model of
the signaling module (see panel G). In C, pComA is held  constant at the LNA mean of 1000 molecules. Progress to competence assumes
differentiation proceeds with time-varying rate proportional to the
level of ComK (see SI), with progress equal to 1 corresponding
to entry to competence. At time zero, the level of ComS and
ComK mRNAs and proteins set to zero. {\bf{(E)}} For the wild-type and SynDM
networks, the percentage change in the fraction of a population of
1000 quorum sensing cells entering competence (within 20 hours) compared
to the fraction when pComA is constant at 1000 molecules, as a function
of the lifetime and CV of the OU input modeling pComA fluctuations.
The limit with pComA constant in each cell is also shown, drawn from a
Gaussian distribution with mean 1000 and the indicated CV.  {\bf{(F)}} For the wild-type and SynDM networks,
the estimated $\mathrm{Prob}[\mathrm{{Competence}}| \langle \mathrm{pComA}\rangle]$
as a function of $\langle \mathrm{pComA}\rangle$, the time-averaged
level of pComA over the 20h experiment, for different OU inputs modeling pComA fluctuations. Estimation
performed using logistic regression. {\bf{(G)}} For the wild-type and SynDM
networks, the fraction of a population of 1000 quorum sensing cells
entering competence as a function of the proportionality constant
of ComK-driven differentiation, for different OU inputs modeling pComA fluctuations (lifetime of 28s corresponds to model of upstream signaling lacking gene expression of the component proteins).
{\bf{(H)}} The autocorrelation function of pComA given by the LNA model of upstream signaling, compared to that of a single OU input process and 2 summed, independent OU processes, both having the mean and variance of pComA given by the LNA. }
\end{center}
\end{figure*}

\section*{Methods}

\subsection*{Validity of the Extrande approach}
The Extrande approach
relies on augmenting the reaction network with an extra, `virtual' channel (giving the augmented system, $Z$), so as to make simulation of the augmented system feasible, while ensuring that the simulated timings and types of biochemical reactions are unaffected by the firings of the extra channel.
In the Extrande method, the conditional
propensity of the extra channel depends on the history of the
extra channel (as well as on the history of the original system, $\mathcal{{H}}_{t}^{X}$), and so does the upper bound. A related Proposition in \cite{OGATA:1981} does not allow for this dependence (see SI).
We therefore provide the new proof below. To see the dependence on the extra channel, note that the bound
is in general updated in {\bk Step 3} of the Extrande algorithm (Box 1) after  
each firing of the extra channel.

The reaction
network to be simulated (Box 1) has the number of molecules of each
species at time $t$ given by
\[
X(t)=X(0)+SR(t),
\]
where $R(t)=\{R_{1}(t),...,R_{M}(t)\}$ is the vector of processes
counting the number of times each biochemical reaction channel fires
during the time interval $[0,t]$, and $S=\{v_{1},...,v_{M}\}$ is
the stoichiometric matrix. The `Poisson' or random time-change representation \cite{Anderson2011} expresses
$R(t)$ in terms of $M$ independent, unit rate Poisson processes,
$Y(t)=\{Y_{1}(t),...,Y_{M}(t)\}$, and so can be written here as
\begin{equation}
\begin{split}
&X(t)=X(0)+\\& S\left[Y_{1}\left(\int_{0}^{t}a_{1}[X(s),I(s)]ds\right),...,Y_{M}\left(\int_{0}^{t}a_{M}[X(s),I(s)]ds\right)\right]^{\mathrm{T}},\label{eq:RTCSys}
\end{split}
\end{equation}
where $I$ is the possibly multivariate input,
superscript $\mathrm{T}$ denotes transpose of a vector, and $a_{j}[X(s),I(s)]$
is the propensity of the $j$th reaction, for $j=1,...,M$, conditional on $\{\mathcal{{H}}_{s}^{X},\mathcal{{I}}\}$. We denote by $\mathcal{{I}}$ the ($\sigma$-field generated by) the entire trajectory of the input.

We introduce as a simulation device the extra, virtual reaction
$R_{M+1}$ : $\emptyset\rightarrow\emptyset$, to form the \textit{augmented
system} 
\[
Z(t)=\left(\begin{array}{c}
X(t)\\
R_{M+1}(t)
\end{array}\right)=\left(\begin{array}{c}
X(0)\\
0
\end{array}\right)+\left(\begin{array}{cc}
S & 0\\
0 & 1
\end{array}\right)\left(\begin{array}{c}
R(t)\\
R_{M+1}(t)
\end{array}\right). \]
The random time-change representation of the augmented system is in terms of $(M+1)$ independent, unit rate Poisson processes,
$Y(t)=\{Y_{1}(t),...,Y_{M+1}(t)\}$
\begin{equation}
\begin{split}
 & Z(t)  =   Z(0)+\left(\begin{array}{cc} 
S & 0\\
0 & 1
\end{array}\right)  \times \\ &
\left(\left[...,Y_{j}\left(\int_{0}^{t}a_{j}[X(s),I(s)]ds\right),...\right],Y_{M+1}\left(\int_{0}^{t}a_{M+1}(s)ds\right)\right)^{\mathrm{T}} ,  
\end{split}
\label{eq:RTCAugSys}
\end{equation}
where $a_{M+1}(s)$ is the propensity of the extra reaction channel
(conditional on $\{{H}_{s}^{Z},\mathcal{{I}}\}$), and where we set
$a_{j}[X(s),I(s)]$, for $j=1,...,M$, as the propensity of the $j$th
reaction conditional on $\{\mathcal{{H}}_{s}^{Z},\mathcal{{I}}\}$,
which now includes the history of the extra channel, $R_{M+1}$.

Notice that Eq. \ref{eq:RTCAugSys} is identical to Eq. \ref{eq:RTCSys}
in its expression of the original system, $X(t)$,
or equivalently of $R(t)$. Therefore, if the propensity $a_{M+1}$
is chosen to somehow make simulation of $[R(t),R_{M+1}(t)]$ straightforward,
we are able to simulate our target, $R(t)$,
by simulating the augmented system in Eq. \ref{eq:RTCAugSys} and
then ignoring $R_{M+1}(t)$. To do this, let $B(t)$ be an $(\mathcal{{H}}_{t}^{Z},\mathcal{{I}})$-measurable
random variable satisfying (with probability 1) that 
\[
a_{0}(t)=\sum_{j=1}^{M}a_{j}[X(t),I(t)]\leq B(t),\text{{} }t\geq0,
\]
so that $B(t)$ is a stochastic upper bound for the total biochemical
reaction propensity. Now define the propensity of the extra channel
(conditional on $\{{H}_{t}^{Z},\mathcal{{I}}\}$) as: 
\[
a_{M+1}(t)=B(t)-a_{0}(t).
\]

The ground process (see SI) of $[R(t),R_{M+1}(t)]$ has propensity (conditional
on $\{\mathcal{{H}}_{t}^{Z},\mathcal{{I}}\}$) given by $\sum_{j=1}^{M+1}a_{j}(t)=B(t),$
by construction. The Extrande method chooses the stochastic bound,
$B(t)$, so that it is constant between firings of the augmented system (see Box 1), which makes straightforward
the simulation of the ground process of \textbf{$[R(t),R_{M+1}(t)]$}.
We write the $i$th occurrence time of the ground process of $[R(t),R_{M+1}(t)]$
as $T_{i},$ $i=1,2,...$ It is now the case that
\begin{equation*}
\mathrm{Prob}\{T_{i+1}-T_{i} \le t|T_{1},Z_{1},...,T_{i},Z_{i},\mathcal{I}\}=1-\exp \{-tB(T_i) \}, 
\end{equation*}
where $Z_{i}$ is
the channel corresponding to the $i$th firing.
The waiting time has an exponential distribution and the occurrence times $\{T_{1},T_{2},...\}$
are therefore just those of a $(\mathcal{{H}}_{t}^{Z},\mathcal{{I}})$-Poisson
process with propensity $B(t)$, and can be simulated analogously
to the SSA as in {\bk Step 4} of Box 1.

What remains is to assign each firing time
$T_{i}$ to one of the $(M+1)$ channels of the augmented system.
We do the allocation sequentially, using the result from counting
process theory \cite[p.34]{Bremaud:81} that, for $j=1,...,(M+1)$:
\begin{equation}
\mathrm{Prob}\{Z_{i+1}=j|T_{1},Z_{1},...,T_{i},Z_{i},T_{i+1},\mathcal{I}\}=\frac{a_{j}[\tilde{X}(T_{i+1}),\tilde{I}(T_{i+1})]}{B(T_{i})},\label{eq:MarkDn-1}
\end{equation}
 where we have used the left-continuous versions $(\tilde{X}(t),\tilde{I}(t))$ of $(X(t),I(t))$, and $\tilde{B}(T_{i+1})=B(T_{i})$. Eq.~\ref{eq:MarkDn-1}
is implemented by {\bk Steps 9-15} in Box 1. The intuition for Eq.~\ref{eq:MarkDn-1}
uses Bayes' theorem. Consider a small interval of time $dt$. The
probability that the channel is the $j$th one given that some reaction
fires at time $T_{i+1}$, since the probability of more than
one reaction can be neglected, is given by $[dt\cdot a_{j}(\tilde{X}_{T_{i+1}},\tilde{I}_{T_{i+1}})]/[dt\cdot{}_{k=1}^{M+1}\sum a_{k}(\tilde{X}_{T_{i+1}},\tilde{I}_{T_{i+1}})].$ 
The target of the Extrande simulation, $R(t)$, is now obtained
by ignoring all the firing times of the extra channel after simulation
of the augmented system is complete. This completes the proof. $\blacksquare$

We note that the condition $\lim_{t\rightarrow\infty}R_{j}(t)=\infty$ $(j=1,...,M)$ is needed for the representation in Eq. \ref{eq:RTCSys}, but is not needed for the validity of the Extrande method. The random time-change representation is used here to make the proof more accessible. The Extrande algorithm results in a probability law, $P$, under which the functions $a_{j}[X(t),I(t)]$ give the propensities of the biochemical reactions conditional upon $(\mathcal{{H}}_{t}^{Z},\mathcal{{I}})$. Because the $a_{j}[X(t),I(t)]$ are $(\mathcal{{H}}_{t}^{X},\mathcal{{I}})$-measurable, they also give the $(\mathcal{{H}}_{t}^{X},\mathcal{{I}})$-conditional propensities of the biochemical reactions under $P$, as required of the probability measure $P$ resulting from the Extrande algorithm.

{\bk Finally, we remark that a description equivalent to the random time-change representation, Eq.~\ref{eq:RTCSys}, is the Chemical Master Equation \cite{Anderson2011}. Specifically, for the conditional probability $P(n,t)=\mathrm{Prob}(X(t)=n|X(0)=n_0;\mathcal{I})$ one can write 
\begin{equation}
 \frac{\mathrm{d} P(n,t)}{\mathrm{d} t} = \sum_{j=1}^M \biggl[ a_{j}[n-v_j,I(t)] P(n-v_j,t) - a_{j}[n,I(t)] P(n,t) \biggr],
\end{equation}
whose propensities are time-varying, stochastic functions due to the dependence on the input process.}

\section*{Acknowledgments}
MV acknowledges support under an MRC Biomedical Informatics Fellowship.


%
%
%
\bibliography{ExtrandeSim}

\begin{thebibliography}{10}
\providecommand{\url}[1]{\texttt{#1}}
\providecommand{\urlprefix}{URL }
\expandafter\ifx\csname urlstyle\endcsname\relax
  \providecommand{\doi}[1]{doi:\discretionary{}{}{}#1}\else
  \providecommand{\doi}{doi:\discretionary{}{}{}\begingroup
  \urlstyle{rm}\Url}\fi
\providecommand{\bibAnnoteFile}[1]{%
  \IfFileExists{#1}{\begin{quotation}\noindent\textsc{Key:} #1\\
  \textsc{Annotation:}\ \input{#1}\end{quotation}}{}}
\providecommand{\bibAnnote}[2]{%
  \begin{quotation}\noindent\textsc{Key:} #1\\
  \textsc{Annotation:}\ #2\end{quotation}}
\providecommand{\eprint}[2][]{\url{#2}}

\bibitem{Covert:2012}
Karr JR, Sanghvi JC, Macklin DN, Gutschow MV, Jacobs JM, et~al. (2012) A
  whole-cell computational model predicts phenotype from genotype.
\newblock Cell 150: 389--401.
\bibAnnoteFile{Covert:2012}

\bibitem{Eldar:2010kk}
Eldar A, Elowitz MB (2010) {Functional roles for noise in genetic circuits}.
\newblock Nature 467: 167--173.
\bibAnnoteFile{Eldar:2010kk}

\bibitem{Thomas2014}
Thomas P, Popovi{\'c} N, Grima R (2014) Phenotypic switching in gene regulatory
  networks.
\newblock Proc Natl Acad Sci USA 111: 6994--6999.
\bibAnnoteFile{Thomas2014}

\bibitem{Crampin:2004}
Crampin EJ, Halstead M, Hunter P, Nielsen P, Noble D, et~al. (2004)
  {Computational physiology and the physiome project}.
\newblock Exp Physiol 89: 1--26.
\bibAnnoteFile{Crampin:2004}

\bibitem{Rand:2012}
Rand U, Rinas M, Schwerk J, Nohren GN, Linnes M, et~al. (2012) {Multi-layered
  stochasticity and paracrine signal propagation shape the type-I interferon
  response}.
\newblock Mol Syst Biol 8: 1--13.
\bibAnnoteFile{Rand:2012}

\bibitem{Hartwell:1999ef}
Hartwell LH, Hopfield JJ, Leibler S, Murray AW (1999) {From molecular to
  modular cell biology}.
\newblock Nature 402: C47--52.
\bibAnnoteFile{Hartwell:1999ef}

\bibitem{Sontag:2007}
Sontag ED (2007) {Monotone and near-monotone biochemical networks}.
\newblock Syst Synth Biol 1: 59--87.
\bibAnnoteFile{Sontag:2007}

\bibitem{Thomas2012}
Thomas P, Straube AV, Grima R (2012) The slow-scale linear noise approximation:
  an accurate, reduced stochastic description of biochemical networks under
  timescale separation conditions.
\newblock BMC Syst Biol 6: 39.
\bibAnnoteFile{Thomas2012}

\bibitem{Bialek2013}
Bialek W (2013) Biophysics: searching for principles.
\newblock Princeton, New Jersey: Princeton University Press.
\bibAnnoteFile{Bialek2013}

\bibitem{Bowsher2013}
Bowsher CG, Voliotis M, Swain PS (2013) {The fidelity of dynamic signaling by
  noisy biomolecular networks}.
\newblock PLoS Comput Biol 9: e1002965.
\bibAnnoteFile{Bowsher2013}

\bibitem{Bowsher2014b}
Bowsher CG, Swain PS (2014) Environmental sensing, information transfer, and
  cellular decision-making.
\newblock Curr Opin Biotechnol 28: 149--155.
\bibAnnoteFile{Bowsher2014b}

\bibitem{Zechner2014}
Zechner C, Koeppl H (2014) Uncoupled analysis of stochastic reaction networks
  in fluctuating environments.
\newblock PLoS Comput Biol 10: e1003942.
\bibAnnoteFile{Zechner2014}

\bibitem{Gillespie:1976}
Gillespie DT (1976) A general method for numerically simulating the stochastic
  time evolution of coupled chemical reactions.
\newblock J Comp Phys 22: 403--434.
\bibAnnoteFile{Gillespie:1976}

\bibitem{Gillespie:1977ww}
Gillespie DT (1977) {Exact stochastic simulation of coupled chemical
  reactions}.
\newblock J Phys Chem 81: 2340--2361.
\bibAnnoteFile{Gillespie:1977ww}

\bibitem{Swain:2002kn}
Swain PS, Elowitz MB, Siggia ED (2002) {Intrinsic and extrinsic contributions
  to stochasticity in gene expression}.
\newblock Proc Natl Acad Sci USA 99: 12795--12800.
\bibAnnoteFile{Swain:2002kn}

\bibitem{Hilfinger:2011ev}
Hilfinger A, Paulsson J (2011) {Separating intrinsic from extrinsic
  fluctuations in dynamic biological systems}.
\newblock Proc Natl Acad Sci USA 108: 12167--12172.
\bibAnnoteFile{Hilfinger:2011ev}

\bibitem{Bowsher2012}
Bowsher CG, Swain PS (2012) {Identifying sources of variation and the flow of
  information in biochemical networks}.
\newblock Proc Natl Acad Sci USA 109: E1320--E1328.
\bibAnnoteFile{Bowsher2012}

\bibitem{Elowitz2002}
Elowitz MB, Levine AJ, Siggia ED, Swain PS (2002) {Stochastic gene expression
  in a single cell}.
\newblock Science 297: 1183--1186.
\bibAnnoteFile{Elowitz2002}

\bibitem{Raser:2004gh}
Raser JM, O'Shea EK (2004) {Control of stochasticity in eukaryotic gene
  expression}.
\newblock Science 304: 1811--1814.
\bibAnnoteFile{Raser:2004gh}

\bibitem{Rosenfeld:2005hn}
Rosenfeld N, Young JW, Alon U, Swain PS, Elowitz MB (2005) {Gene regulation at
  the single-cell level}.
\newblock Science 307: 1962--1965.
\bibAnnoteFile{Rosenfeld:2005hn}

\bibitem{Zechner:2014}
Zechner C, Unger M, Pelet S, Peter M, Koeppl H (2014) {Scalable inference of
  heterogeneous reaction kinetics from pooled single-cell recordings}.
\newblock Nat Methods 11: 197.
\bibAnnoteFile{Zechner:2014}

\bibitem{DasNeves:2010km}
Das~Neves RP, Jones NS, Andreu L, Gupta R, Enver T, et~al. (2010) {Connecting
  variability in global transcription rate to mitochondrial variability}.
\newblock PLoS Biol 8: e1000560.
\bibAnnoteFile{DasNeves:2010km}

\bibitem{Eser:2014}
Eser P, Demel C, Maier KC, Schwalb B, Pirkl N, et~al. (2014) {Periodic mRNA
  synthesis and degradation co-operate during cell cycle gene expression}.
\newblock Mol Syst Biol 10: 717--717.
\bibAnnoteFile{Eser:2014}

\bibitem{Levine:2013}
Levine JH, Lin Y, Elowitz MB (2013) {Functional roles of pulsing in genetic
  circuits}.
\newblock Science 342: 1193--1200.
\bibAnnoteFile{Levine:2013}

\bibitem{Cheong:2011jp}
Cheong R, Rhee A, Wang CJ, Nemenman I, Levchenko A (2011) {Information
  transduction capacity of noisy biochemical signaling networks}.
\newblock Science 334: 354--358.
\bibAnnoteFile{Cheong:2011jp}

\bibitem{Bowsher2014}
Voliotis M, Perrett RM, McWilliams C, McArdle CA, Bowsher CG (2014) Information
  transfer by leaky, heterogeneous, protein kinase signaling systems.
\newblock Proc Natl Acad Sci USA 111: E326--E333.
\bibAnnoteFile{Bowsher2014}

\bibitem{Guerriero:2012}
Guerriero ML, Pokhilko A, Fernandez AP, Halliday KJ, Millar AJ, et~al. (2012)
  {Stochastic properties of the plant circadian clock}.
\newblock J R Soc Interface 9: 744--756.
\bibAnnoteFile{Guerriero:2012}

\bibitem{Anderson:2007}
Anderson DF (2007) {A modified next reaction method for simulating chemical
  systems with time dependent propensities and delays}.
\newblock J Chem Phys 127: 214107.
\bibAnnoteFile{Anderson:2007}

\bibitem{Shahrezaei:2008iq}
Shahrezaei V, Ollivier JF, Swain PS (2008) {Colored extrinsic fluctuations and
  stochastic gene expression}.
\newblock Mol Syst Biol 4: 196.
\bibAnnoteFile{Shahrezaei:2008iq}

\bibitem{Johnston:2012}
Johnston IG, Gaal B, Neves RPd, Enver T, Iborra FJ, et~al. (2012)
  {Mitochondrial variability as a source of extrinsic cellular noise}.
\newblock PLoS Comput Biol 8: e1002416.
\bibAnnoteFile{Johnston:2012}

\bibitem{lewis1979}
Lewis PA, Shedler GS (1979) Simulation of nonhomogeneous poisson processes by
  thinning.
\newblock Naval Research Logistics Quarterly 26: 403--413.
\bibAnnoteFile{lewis1979}

\bibitem{OGATA:1981}
Ogata Y (1981) {On Lewis Simulation Method for point-processes}.
\newblock IEEE Trans Inf Theory 27: 23--31.
\bibAnnoteFile{OGATA:1981}

\bibitem{Thanh2015}
Thanh VH, Priami C (2015) Simulation of biochemical reactions with
  time-dependent rates by the rejection-based algorithm.
\newblock J Chem Phys 143: 054104.
\bibAnnoteFile{Thanh2015}

\bibitem{higham2001}
Higham DJ (2001) An algorithmic introduction to numerical simulation of
  stochastic differential equations.
\newblock SIAM review 43: 525--546.
\bibAnnoteFile{higham2001}

\bibitem{DelVecchio:2008gy}
Del~Vecchio D, Ninfa AJ, Sontag ED (2008) {Modular cell biology: retroactivity
  and insulation}.
\newblock Mol Syst Biol 4: 161.
\bibAnnoteFile{DelVecchio:2008gy}

\bibitem{panda2002}
Panda S, Antoch MP, Miller BH, Su AI, Schook AB, et~al. (2002) Coordinated
  transcription of key pathways in the mouse by the circadian clock.
\newblock Cell 109: 307--320.
\bibAnnoteFile{panda2002}

\bibitem{Millar:2011}
O'Neill JS, van Ooijen G, Dixon LE, Troein C, Corellou F, et~al. (2011)
  {Circadian rhythms persist without transcription in a eukaryote}.
\newblock Nature 469: 554--558.
\bibAnnoteFile{Millar:2011}

\bibitem{Thomas:2012}
Thomas P, Matuschek H, Grima R (2012) Intrinsic noise analyzer: a software
  package for the exploration of stochastic biochemical kinetics using the
  system size expansion.
\newblock PLoS one 7: e38518.
\bibAnnoteFile{Thomas:2012}

\bibitem{Maamar:2007}
Maamar H, Raj A, Dubnau D (2007) {Noise in gene expression determines cell fate
  in Bacillus subtilis}.
\newblock Science 317: 526--529.
\bibAnnoteFile{Maamar:2007}

\bibitem{Suel:2006ea}
S{\"u}el GM, Garcia-Ojalvo J, Liberman LM, Elowitz MB (2006) {An excitable gene
  regulatory circuit induces transient cellular differentiation}.
\newblock Nature 440: 545--550.
\bibAnnoteFile{Suel:2006ea}

\bibitem{Suel:2007gp}
S{\"u}el GM, Kulkarni RP, Dworkin J, Garcia-Ojalvo J, Elowitz MB (2007)
  {Tunability and noise dependence in differentiation dynamics}.
\newblock Science 315: 1716--1719.
\bibAnnoteFile{Suel:2007gp}

\bibitem{Cagatay:2009ie}
Ca{\u g}atay T, Turcotte M, Elowitz MB, Garcia-Ojalvo J, S{\"u}el GM (2009)
  {Architecture-dependent noise discriminates functionally analogous
  differentiation circuits}.
\newblock Cell 139: 512--522.
\bibAnnoteFile{Cagatay:2009ie}

\bibitem{Kampen:1961}
Kampen NG (1961) {A power series expansion of the master equation}.
\newblock Can J Phys 39: 551--567.
\bibAnnoteFile{Kampen:1961}

\bibitem{Michelsen:2003}
Michelsen O, de~Mattos MJT, Jensen PR, Hansen FG (2003) {Precise determinations
  of C and D periods by flow cytometry in Escherichia coli K-12 and B/r}.
\newblock Microbiology 149: 1001--1010.
\bibAnnoteFile{Michelsen:2003}

\bibitem{Kuchina:2011}
Kuchina A, Espinar L, Cagatay T, Balbin AO, Zhang F, et~al. (2011) {Temporal
  competition between differentiation programs determines cell fate choice}.
\newblock Mol Syst Biol 7.
\bibAnnoteFile{Kuchina:2011}

\bibitem{Hahn:2005}
Hahn J, Maier B, Haijema BJ, Sheetz M, Dubnau D (2005) {Transformation proteins
  and DNA uptake localize to the cell poles in Bacillus subtilis}.
\newblock Cell 122: 59--71.
\bibAnnoteFile{Hahn:2005}

\bibitem{Anderson2011}
Anderson DF, Kurtz TG (2011) Continuous time markov chain models for chemical
  reaction networks.
\newblock In: Design and analysis of biomolecular circuits, Springer. pp.
  3--42.
\bibAnnoteFile{Anderson2011}

\bibitem{Bremaud:81}
Br{\'e}maud P (1981) Point processes and queues.
\newblock Berlin, Germany: Springer.
\bibAnnoteFile{Bremaud:81}

\end{thebibliography}

%

%
%
%

\section*{Supporting Information Legends}
%

%
\begin{description}
\item{\bf SI. Supporting Information: Stochastic Simulation of Biomolecular Networks in Dynamic Environments}. 
\end{description}

\end{document}